\RequirePackage{fix-cm}
\documentclass[twocolumn,epjc3]{svjour3}  
\smartqed  % flush right qed marks, e.g. at end of proof
\RequirePackage{graphicx}
\RequirePackage{amsmath}
\RequirePackage{multicol}
\journalname{Eur. Phys. J. C}
\begin{document}

\title{Some remarks on Hayward black hole with a cloud of strings
}
%\subtitle{Hayward black hole with a cloud of strings}


\author{F. F. Nascimento\thanksref{e1, addr1}
        \and
        V. B. Bezerra\thanksref{e2,addr1} 
        \and J. M. Toledo\thanksref{e3,addr1}
}

\thankstext{e1}{e-mail: fran.nice.fisica@gmail.com}
\thankstext{e2}{e-mail: valdir@fisica.ufpb.br}
\thankstext{e3}{e-mail: jefferson.m.toledo@gmail.com}


\institute{Departamento de Física, 
  Universidade Federal da Paraíba, Caixa Postal 5008, 58059-900, João Pessoa, PB, Brazil \label{addr1}
}

\date{Received: date / Accepted: date}

\maketitle

\begin{abstract}
We obtain the metric corresponding to the Hayward black hole space-time surrounded by a cloud of strings and investigate the role played by this cloud on the horizons, geodesics, effective potential and thermodynamics. We compare the obtained results with the ones of the literature, corresponding to the Hayward black hole, when the cloud of strings is absent. Also, the question related to its nature, with respect to regularity, in this scenario, is examined.
\end{abstract}

\section{Introduction}
\label{intro}

Black hole solutions of Einstein's equations were known since the middle 1910s, and then shortly after the formulation of the General Theory of Relativity. The simplest vacuum solution, which describes the gravitational field of a static, uncharged, and spherically symmetric body, was obtained by Schwarzschild \cite{schwarzschild1916uber}. Its generalization, to include the
presence of electric charge as a source was obtained by Reissner and Nordström \cite{reissner1916eigengravitation,nordstrom1918een}. After nearly fifty years, Kerr \cite{kerr1963gravitational} obtained a generalization of the Schwarzschild solution, by considering rotation and it took only two years before the metric of a charged and rotating gravitational body to be obtained, the well-known Kerr-Newman solution \cite{newman1965metric}. It is worth emphasizing that the metrics corresponding to these black hole solutions have a curvature singularity at $r = 0$, whose existence creates some difficulties in the General Theory of Relativity because, at the singularity point, the physical quantities diverge and, therefore, the physical laws are not valid.

To solve these difficulties related to the curvature singularity and its
consequences, some black hole solutions have been considered, whose
metrics and curvature invariants have no singularity, that is, they are
regular everywhere, particularly, at the origin. These solutions correspond to what are called Regular
Black Holes. The first solution of a static spherically symmetric Regular
Black Hole dated back to the later 1960s and was obtained by Bardeen \cite{bardeen1968non}.
Nowadays, there are several different static Regular Black Hole solutions,
among them, we can mention \cite{dymnikova1992vacuum,dymnikova2003spherically,ayon1998regular,ayon1999new,mars1996models,sajadi2017nonlinear,balart2014regular,frolov2016notes,bronnikov2001regular}, in special, the one obtained by Hayward \cite{hayward2006formation}. The solution of the field equations presented by Hayward is free of charge term and its physical aspects are quite like Bardeen's solution \cite{bardeen1968non}.
The Hayward black hole solution \cite{hayward2006formation} is static, uncharged and spherically symmetric. It is worth calling attention to the fact that this solution becomes a de Sitter space-time at the center of the black hole, and therefore, there is no singularity at $r=0$, and, additionally, it is asymptotically flat as $r \rightarrow \infty$.

Originally, the Hayward black hole solution was obtained from the modified Einstein equations, with the parameter appearing in the solution being related to the energy level in the near-horizon region of the black hole \cite{hayward2006formation}, which can be viewed as a constant acting in this space-time. On the other hand, the Hayward black hole solution can also be obtained in the context of a gravity theory coupled with nonlinear electrodynamics, in which case, the parameter mentioned is no more a universal constant, but a magnetic charge.

The Regular Black Hole solutions and the interesting consequences arising from these solutions have inspired further investigations related to such black holes, as, for example, those regarding particle geodesics \cite{abbas2014geodesic,chiba2017note,zhou2012geodesic,wei2015null}, structure and lens effect \cite{perez2018region,eiroa2011gravitational}, thermodynamics \cite{saleh2018thermodynamics,rodrigue2020thermodynamic,molina2021thermodynamics,iguchi2023gravitational} and quasi-normal modes \cite{fernando2012quasinormal,flachi2013quasinormal,lin2013quasinormal}. 

In the later 1970s, Letelier \cite{letelier1979clouds} obtained general solutions of the Einstein
equations corresponding to spherically, plane-symmetric and cylindrically symmetric space-times, by considering a cloud of strings as sources of the gravitational field. In the first case, namely, when a spherically symmetric cloud of strings, radially directed, surrounds the gravitating body, the obtained solution is basically the Schwarzschild black hole solution
slightly modified, in such a way that the metric is similar to the Schwarzschild one, but with a solid deficit angle which depends on the parameter associated with the presence of the cloud of strings. Therefore, the gravitational effects are of global origin, with respect to the cloud. As an example of these effects, we mention the fact that the radius of the event horizon is enlarged as compared with the Schwarzschild radius.

Given the possible astrophysical consequences, it is important to investigate the gravitational consequences when a black hole is immersed in a cloud of strings. With this aim, several studies concerning different aspects associated with the physics of black holes surrounded by a cloud of strings were performed during the last decades, in the context of the General Theory of Relativity \cite{batool2017null,sood2022thermodynamic,rodrigues2022bardeen,gracca2018effects,rodrigues2022embedding,ghaffarnejad2019last}, as well as in different modified versions of this Theory \cite{morais2018cloud,toledo2019black,ghaffarnejad2018effects,herscovich2010black,ghosh2014cloud,cai2020quasinormal}. Also, in which concerns the Regular Black Holes, some studies have been done by considering the presence of a cloud of strings\cite{rodrigues2022bardeen,atamurotov2023weak}, and some aspects of the thermodynamics were investigated, with emphasis to the role played by the cloud of strings.

In this paper, we investigate the role played by a cloud of strings surrounding a Hayward black hole on the horizons, singular behavior, geodesics, effective potential, and thermodynamics as compared to the case where the cloud is absent. Also, some discussion about the regularity is presented.

The paper is organized as follows. In Sect. \ref{sec2}, we review the Hayward black hole solution and obtain the Hayward black hole solution in the case in which this gravitational body is surrounded by a cloud of strings. In Sect. \ref{sec3}, we focus the discussion on the horizons,
singularity, geodesics, and effective potential. Section \ref{sec4} is devoted to different aspects of thermodynamics, with emphasis on the role played by the parameter that codifies the presence of the cloud of strings. In Section \ref{sec5}, we briefly summarize our results.

\section{The metric}
\label{sec2}

In this section, we obtain the metric corresponding to the Hayward black hole with a cloud of strings and analyze its properties.

\subsection{The Hayward solution}

The metric of the non-singular(regular) black hole obtained by Hayward \cite{hayward2006formation} is given by

\begin{equation} \label{hayward}
    ds^2 = f(r) dt^2 - f(r)^{-1} dr^2 - r^2 d\Omega^2,
\end{equation}

\noindent where $l$ and $m$ are positive constants, and $d \Omega^2 = d\theta^2 + \sin^2 \theta d\phi^2$. The function $f(r)$ is given by

\begin{equation}
    f(r) = 1-\frac{2 m r^2}{r^3 + 2 l^2 m}
\end{equation}

Note that 

\begin{equation}
\lim_{r\rightarrow 0}\left(1-\frac{2mr^2}{r^3+2l^2m}\right)=1,
\label{eq:2.40}
\end{equation}

\noindent which means that the Hayward metric is a regular (non-singular) solution of Einstein equations. Furthermore, we can observe this property by calculating the Kretschmann scalar, which is given by

\begin{equation}
\begin{aligned}
K &=R_{\alpha\beta\mu\nu}R^{\alpha\beta\mu\nu}\\
&=\frac{48m^2(r^{12}-4r^9g^3+18r^6g^6-2r^3g^9+2g^{12})}{(r^3+g^3)^6},
\label{eq:2.40.1}
\end{aligned}
\end{equation}

In the limit $r \rightarrow 0$, we get

\begin{equation}
\begin{aligned}
\lim_{r\rightarrow 0} K&=\lim_{r\rightarrow 0}\frac{48m^2(r^{12}-4r^9g^3+18r^6g^6-2r^3g^9+2g^{12})}{(r^3+g^3)^6}\\
&=\frac{96m^2}{g^6}.
\end{aligned}
\label{eq:2.40.2}
\end{equation}

Using the metric given by Eq. (\ref{hayward}), we can obtain the following components of the Einstein tensor \cite{hayward2006formation}:

\begin{equation}
G_t^{\;t}=G_r^{\;r}=\frac{12l^2m^2}{(r^3+g^3)^2},
\label{eq:1.18}
\end{equation}
\begin{equation}
G_\theta^{\;\theta}=G_\phi^{\;\phi}=-\frac{24(r^3-l^2m)l^2m^2}{(r^3+g^3)^3},
\label{eq:1.19}
\end{equation}

 \noindent where
 
\begin{equation}
g^3\equiv 2l^2m.
\label{eq:1.20}
\end{equation}

\noindent which, through Einstein equations, are proportional to the stress-energy tensor of the source.

In Hayward's solution, the function $f(r)$ can also be written as

\begin{equation}
f(r)=1-\frac{2m(r)}{r},
\label{eq:2.41}
\end{equation}

\noindent where 

\begin{equation}
m(r)=\frac{mr^3}{(r^3+g^3)},
\label{eq:2.42}
\end{equation}

\noindent is the black hole mass, which depends on the radial coordinate and $g^3$ is given by Eq.(\ref{eq:1.20}). We can observe that, if $r\rightarrow \infty$,  $m(r) \rightarrow m$. Thus, very far from the Black Hole, the Hayward solution is similar to the Schwarzschild one.

For $r \rightarrow 0$ (small values of $r$), we can write 

\begin{equation}
m(r) \approx \frac{mr^3}{g^3},
\label{eq:2.43}
\end{equation}

\noindent and then

\begin{equation}
f(r)\approx 1-Cr^2,
\label{eq:2.44}
\end{equation}

\noindent with $C=\frac{2m}{g^3}$ being a positive constant. Observe that the space-time metric with $f(r)$ given by Eq. (\ref{eq:2.44}) is similar to De Sitter space-time. Thus, the Hayward black hole has an internal core with behavior similar to de Sitter metric \cite{hayward2006formation}.
%%%%%%%%%%%%%%%%%%%%%%%%%%%%%%%%%%%%%%%%%%%%%%%%%%%%%%%%%%%%%%%%%%%%%%

\subsection{Hayward black hole with a cloud of strings}

Now, let us consider the Hayward black hole with a cloud of strings.

A spherically symmetric cloud of strings is described by the stress-energy tensor  \cite{letelier1979clouds}

\begin{equation}
T^{\mu\nu}=\rho\frac{\Sigma^{\mu\beta}\Sigma_{\beta}^{\;\nu}}{(-\gamma)^{1/2}},
\label{eq:1.41}
\end{equation}

\noindent where $\Sigma^{\mu\nu}$ is a bivector that represents the world sheet of the strings and is given by

\begin{equation}
\Sigma^{\mu\nu}=\epsilon^{ab}\frac{\partial{x^\mu}}{\partial{\lambda^{a}}}\frac{\partial{x^\nu}}{\partial{\lambda^{b}}},
\label{eq:1.42}
\end{equation}

\noindent where $\epsilon^{ab}$ is the Levi-Civita bidimensional symbol and $\epsilon^{01}=-\epsilon^{10}=1$.

The non-null components of the stress-energy tensor of the cloud of strings are given by \cite{letelier1979clouds}:

\begin{equation}
T_{0}^{\;0}=T_{1}^{\;1}=\frac{a}{r^2},
\label{eq:1.43}
\end{equation}

\begin{equation}
T_{2}^{\;2}=T_{3}^{\;3}=0.
\label{eq:1.44}
\end{equation}

Now, let us consider the line element corresponding to a Hayward black hole with a cloud of strings given by \cite{d2022introducing}:

\begin{equation}
ds^2=e^\nu dt^2-e^\lambda dr^2-r^2 d\theta^2-r^2\sin^2\theta d\phi^2,
\label{eq:1.21}
\end{equation}

\noindent where, $\nu$ and $\lambda$ are functions of the radial coordinate, $r$, only, if we assume that the metric is static.

The non-null components of the Einstein tensor obtained by the metric of Eq. (\ref{eq:1.21}) are given by

\begin{equation}
G_{t}^{\;t}=e^{-\lambda}\left(\frac{\lambda'}{r}-\frac{1}{r^2}\right)+\frac{1}{r^2}.
\label{eq:1.22}
\end{equation}

\begin{equation}
G_{r}^{\;r}=-e^{-\lambda}\left(\frac{\nu'}{r}+\frac{1}{r^2}\right)+\frac{1}{r^2},
\label{eq:1.23}
\end{equation}

\begin{equation}
G_{\theta}^{\;\theta}=G_{\phi}^{\;\phi}=\frac{1}{2}e^{-\lambda}\left(\frac{\nu'\lambda'}{2}+\frac{\lambda'}{r}-\frac{\nu'}{r}-\frac{\nu'^2}{2}-\nu''\right).
\label{eq:1.24}
\end{equation}

Thus, considering that there is no interaction between the cloud of strings and the black hole, the total stress-energy tensor can be obtained by the linear superposition of the individual ones. Thus, using Eqs.(\ref{eq:1.22})-(\ref{eq:1.24}), (\ref{eq:1.18})-(\ref{eq:1.19}) and (\ref{eq:1.43})-(\ref{eq:1.44}) the Einstein equations are given by:

\begin{equation}
e^{-\lambda}\left(\frac{\lambda'}{r}-\frac{1}{r^2}\right)+\frac{1}{r^2}=\frac{12l^2m^2}{(r^3+2l^2m)^2}+\frac{a}{r^2},
\label{eq:1.45}
\end{equation}

\begin{equation}
-e^{-\lambda}\left(\frac{\nu'}{r}+\frac{1}{r^2}\right)+\frac{1}{r^2}=\frac{12l^2m^2}{(r^3+2l^2m)^2}+\frac{a}{r^2},
\label{eq:1.46}
\end{equation}

\begin{equation}
\begin{aligned}
&\frac{1}{2}e^{-\lambda}\left(\frac{\nu'\lambda'}{2}+\frac{\lambda'}{r}-\frac{\nu'}{r}-\frac{\nu'^2}{2}-\nu''\right) =\\
& -\frac{24(r^3-l^2m)l^2m^2}{(r^3+2l^2m)^3}.
\end{aligned}
\label{eq:1.47}
\end{equation}

Subtracting Eqs. (\ref{eq:1.45}) and (\ref{eq:1.46}), we obtain

\begin{equation}
\lambda=-\nu\Rightarrow\lambda'=-\nu'.
\label{eq:1.48}
\end{equation}

Summing Eqs.(\ref{eq:1.45}) and (\ref{eq:1.46}) and taking into account Eq.(\ref{eq:1.48}), we obtain

\begin{equation}
e^{-\lambda}\frac{\lambda'}{r}-e^{-\lambda}\frac{1}{r^2}+\frac{1}{r^2}=\frac{12l^2m^2}{(r^3+2l^2m)^2}+\frac{a}{r^2}.
\label{eq:1.49}
\end{equation}

Now, let us write the following relations 

\begin{equation}
\nu=-\lambda=ln(1+f(r)).
\label{eq:1.30}
\end{equation}

\noindent Taking into account Eqs.(\ref{eq:1.48}) and (\ref{eq:1.30}), we can write Eqs.(\ref{eq:1.49}) and (\ref{eq:1.47}), respectively,  as follows:

\begin{equation}
-\frac{1}{r^2}(rf'+f)=\frac{12l^2m^2}{(r^3+2l^2m)^2}+\frac{a}{r^2},
\label{eq:1.50}
\end{equation}

\begin{equation}
2\frac{f'}{r}+f''=48\frac{(r^3-l^2m)l^2m^2}{(r^3+2l^2m)^3}.
\label{eq:1.51}
\end{equation}

Summing Eqs.(\ref{eq:1.50}) and (\ref{eq:1.51}) and multiplying by $r^2$, we get:

\begin{equation}
r^2f''+rf'-f-a-\frac{12l^2m^2r^2}{(r^3+2l^2m)^2}-48\frac{(r^3-l^2m)l^2m^2r^2}{(r^3+2l^2m)^3}=0,
\label{eq:1.52}
\end{equation}

\noindent whose solution is given by

\begin{equation}
f(r)=\frac{4l^2m^2-2al^2mr-ar^4}{r(r^3+2l^2m)} +\frac{C_1}{r}+rC_2,
\label{eq:1.53}
\end{equation}

\noindent where we can adopt $C_1=-2m$ and $C_2=0$. Substituting Eq.(\ref{eq:1.53}) into
Eq.(\ref{eq:1.30}), we get

\begin{equation}
\nu=-\lambda=ln\left(1-a-\frac{2mr^2}{r^3+2l^2m}\right).
\label{eq:1.54}
\end{equation}

Thus, the metric of the Hayward black hole with a cloud of strings is  written as

\begin{equation} 
\begin{aligned}
ds^2 &=\left(1-a-\frac{2mr^2}{r^3+2l^2m}\right)dt^2\\
&-\left(1-a-\frac{2mr^2}{r^3+2l^2m}\right)^{-1}dr^2 \\
& -r^2 d\theta^2-r^2\sin^2\theta d\phi^2.
\end{aligned}
\label{eq:1.57}
\end{equation}

Note that, if $a=0$, we obtain the Hayward metric\cite{hayward2006formation} given by Eq. (\ref{hayward}).  If $l=0$, the metric of Eq.(\ref{eq:1.57}) becomes similar to the Letelier spacetime \cite{letelier1979clouds}.

Thus, for the metrics given by Eq. (\ref{eq:1.57}), the Kretschmann scalar is 

\begin{equation}
\begin{aligned}
K&=R_{\alpha\beta\mu\nu}R^{\alpha\beta\mu\nu}=\frac{4 a^2}{r^4}+\frac{16a m}{g^3 r^2+r^5}\\
&+\frac{48 m^2 \left(r^{12}-4r^9g^3 +18r^6g^6-2r^3g^9+2g^{12}\right)}{\left(r^3+g^3\right)^6},
\end{aligned}
\label{eq:1.57.1}
\end{equation}

If we neglect the could of strings, $a=0$, we obtain the Kretschmann scalar given in Eq. (\ref{eq:2.40.1}). Now, let us determine the value of these results in the limits $r\rightarrow 0$ and $r\rightarrow \infty$.

\begin{equation}
\begin{aligned}
    \lim_{r\rightarrow 0} K&=\lim_{r\rightarrow 0}\left(\frac{4 a^2}{r^4}+\frac{16a m}{g^3 r^2+r^5} \right.\\
    &\left.+\frac{48 m^2 \left(r^{12}-4r^9g^3 +18r^6g^6-2r^3g^9+2g^{12}\right)}{\left(r^3+g^3\right)^6}\right)=\infty,
\end{aligned}
\label{eq:1.57.2}
\end{equation}

\begin{equation}
\begin{aligned}
\lim_{r\rightarrow \infty} K&=\lim_{r\rightarrow \infty}\left(\frac{4 a^2}{r^4}+\frac{16a m}{g^3 r^2+r^5} \right.\\
&+\left.\frac{48 m^2 \left(r^{12}-4r^9g^3 +18r^6g^6-2r^3g^9+2g^{12}\right)}{\left(r^3+g^3\right)^6}\right)=0. 
\end{aligned}
\label{eq:1.57.3}
\end{equation}

Therefore, from the analysis of the Kretschmann scalar, in the limit 
$ r\rightarrow 0$, we 
conclude that the inclusion of the cloud of strings influences the metric, by destroying the 
regularity, and as a consequence, introducing a singularity. In other words, the metric of the Hayward black hole with a cloud of strings is singular in the origin ($r=0$). 

%%%%%%%%%%%%%%%%%%%%%%%%%%%%%%%%%%%%%%%%%%%%%%%%%%%%%%%%%%%%%%
\section{Black hole horizons, geodesics and effective potential}
\label{sec3}

\subsection{
Black hole horizons}

In what follows, we study the horizons of the Hayward black hole space-time, given by the line element shown in Eq.(\ref{eq:1.57}). From now on, let us identify the function $g(r)$ as

\begin{equation}
g(r)=1-a-\frac{2mr^2}{r^3+2l^2m}
\label{eq:1.58}
\end{equation}

Analyzing the roots of the function $g(r)$, we can obtain a critical black hole mass, which is given by  
\begin{equation}
m_{*}=\frac{3}{4} \sqrt{3} (1-a)^{3/2}l,
\label{eq:1.58.1}
\end{equation}
as well as a critical value for the radial coordinate, $r$, shown in what follows

\begin{equation}
\begin{aligned}
r_{*}&=-\frac{\sqrt{3}}{2 (1-a)^{5/2}} \left[a^3-3 a^2+a \sqrt{1-a} \sqrt[3]{-(1-a)^{9/2}} \right.\\
&\left.+\left(-(1-a)^{9/2}\right)^{2/3}+3 a-\frac{(a-1)^5}{\left(-(1-a)^{9/2}\right)^{2/3}}-1\right] l.
\end{aligned}
\label{eq:2.76}
\end{equation}

\begin{figure*}
\centering
\begin{minipage}[!]{0.45\linewidth}
\includegraphics[scale=0.8]{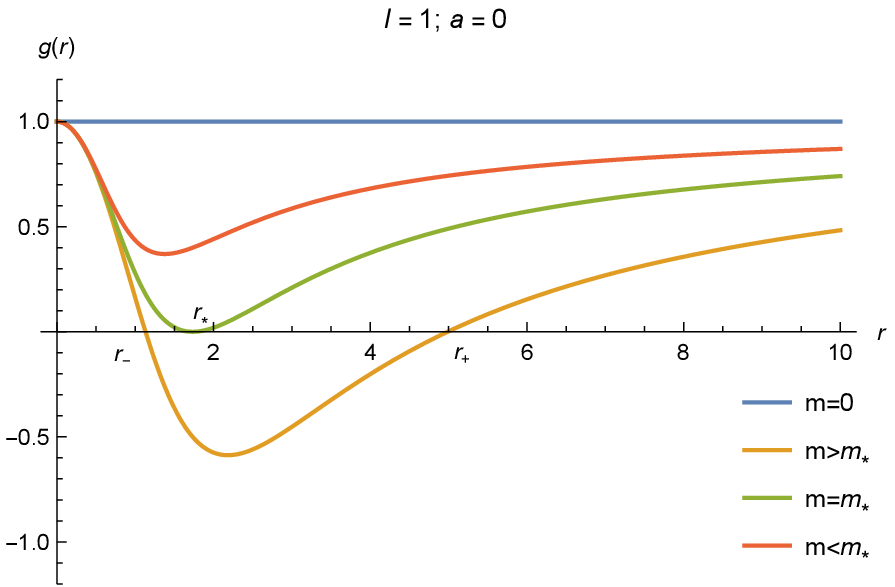}
\end{minipage}
\begin{minipage}[!]{0.45\linewidth}
\includegraphics[scale=0.8]{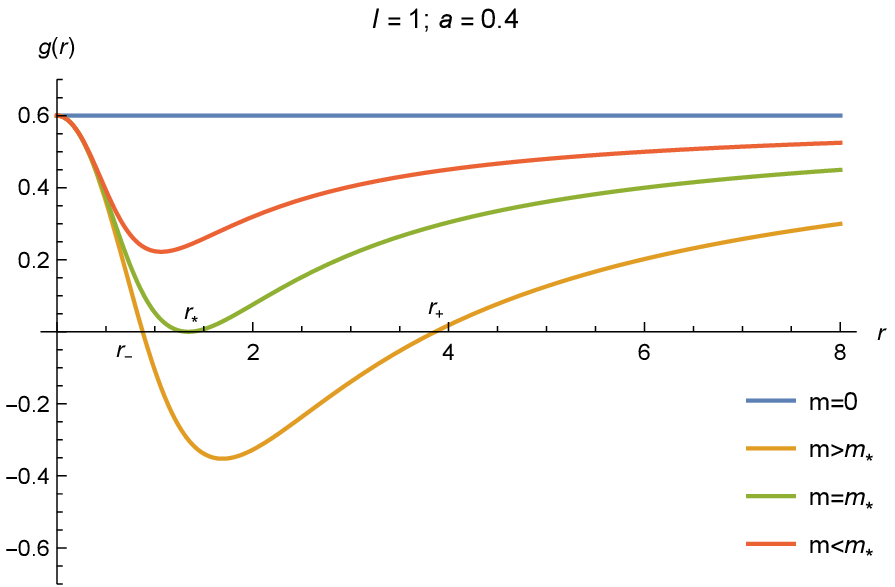}
\end{minipage}
\begin{minipage}[!]{0.45\linewidth}
\includegraphics[scale=0.8]{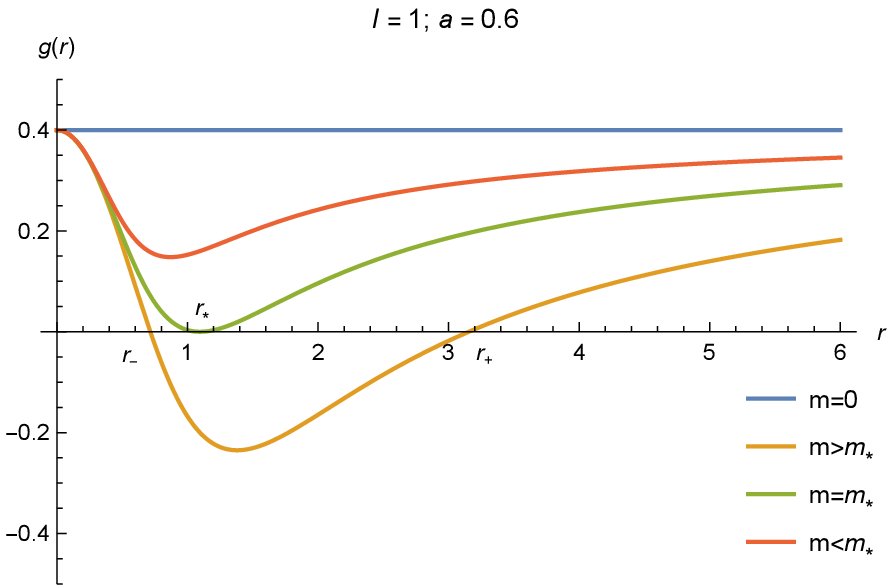}
\end{minipage}
\begin{minipage}[!]{0.45\linewidth}
\includegraphics[scale=0.8]{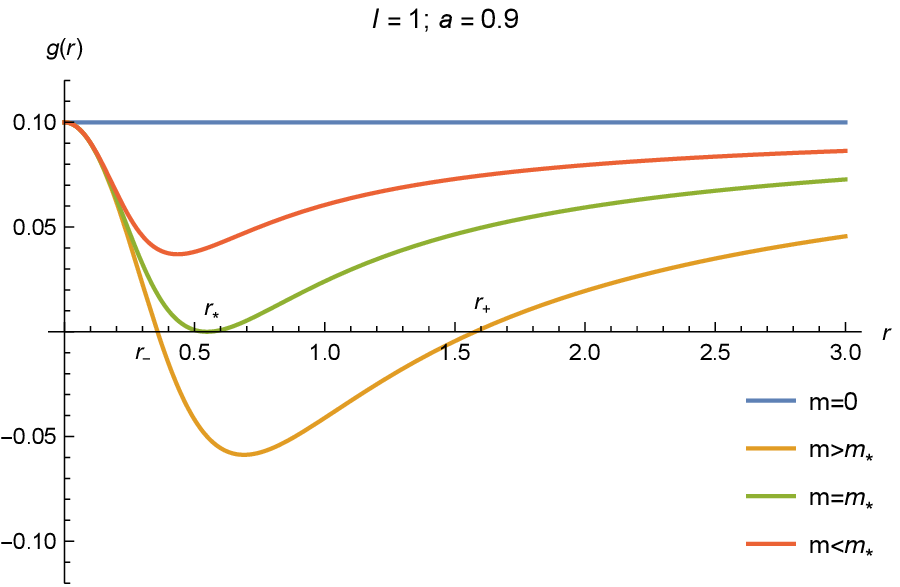}
\end{minipage}
\caption{The function $g(r)$ for $l=1$ and different values of $m$: $m=0$, $m<m_{*}$, $m=m_{*}$ and $m>m_{*}$.} 
\label{im1}
\end{figure*}

Considering only positive values of $r$, if the black hole mass is higher than the critical mass, $m>m_*$, $g(r)$ has two real roots. If $m=m_*$, $g(r)$ has a unique real root, which is equal to $r_*$. Finally, $g(r)$ has no real roots for $m<m_*$. If we neglect the cloud of strings, $a=0$, the critical mass is reduced to $m_*= (3\sqrt{3}/4) l$ and the critical radius is $r_{*}=\sqrt{3}l$, quantities already obtained by Hayward \cite{hayward2006formation}. The described behavior of $g(r)$ can be observed in Fig. \ref{im1}.

\subsection{Black hole geodesics}

Given the space-time metric, the trajectory of particles and light can be described by the geodesic motion. The geodesic equations can also be obtained from the Lagrangian given in the equation

\begin{equation*}
\mathcal{L}=\frac{1}{2}g_{\mu\nu}\frac{dx^\mu}{d\tau}\frac{dx^\nu}{d\tau}=\frac{1}{2} g_{\mu\nu}\dot{x}^\mu\dot{x}^\nu,
\end{equation*}

\noindent which, in the space-time of the Hayward black hole with a cloud of strings, can be written as

\begin{equation}
\begin{aligned}
\mathcal{L}&=\frac{1}{2}\left[\left(1-a-\frac{2mr^2}{r^3+2l^2m}\right)\dot{t}^2-\frac{1}{\left(1-a-\frac{2mr^2}{r^3+2l^2m}\right)}\dot{r}^2 \right.\\&
-r^2\dot{\theta}^2 -\left.r^2\sin^2{\theta}\dot{\phi}^2\right],
\end{aligned}
\label{eq:1.76}
\end{equation}

\noindent where the dot represents the derivative in respect to the proper time $\tau$. Rescaling the parameter $\tau$, we can define   $L=2\mathcal{L}$, which, for time-like geodesics, is equal to $+1$, for space-like geodesics is equal to $-1$ and is equal to $0$ for null geodesics \cite{chandrasekhar1983mathematical}. 

The Euler-Lagrange equations are given by

\begin{equation}
\frac{d}{d\tau}\left(\frac{\partial \mathcal{L}}{\partial\dot{x}^\mu}\right)-\frac{\partial \mathcal{L}}{\partial x^\mu}=0.
\label{eq:1.75}
\end{equation}

For $\mu=0$ and $\mu=3$ in Eq.(\ref{eq:1.75}), with $\mathcal{L}$ given by Eq.(\ref{eq:1.76}), we get, respectively:

\begin{equation}
\dot{t}=\frac{E}{\left(1-a-\frac{2mr^2}{r^3+2l^2m}\right)},
\label{eq:1.77}
\end{equation}

\begin{equation}
\dot{\phi}=-\frac{J}{r^2\sin^2\theta},
\label{eq:1.78}
\end{equation}

\noindent where $E$ and $J$ are movement constants which corresponds to the Killing vectors $\partial_t$ and $\partial_\phi$, respectively. We can interpret these constants as the energy $E$ and the angular momentum $J$ of the particle which is moving nearby the black hole. 

Let us restrict the analysis of the geodesics to the equatorial plane of the black hole, $\theta=\frac{\pi}{2}$. Doing that, the Eqs. (\ref{eq:1.77})-(\ref{eq:1.78}) are reduced to

\begin{equation}
\dot{t}=\frac{E}{\left(1-a-\frac{2mr^2}{r^3+2l^2m}\right)},
\label{eq:1.79}
\end{equation}

\begin{equation}
\dot{\phi}=-\frac{J}{r^2}.
\label{eq:1.80}
\end{equation}

\noindent where $\dot{t}$ and $\dot{\phi}$ are the derivatives of $t$ and $\phi$ with respect to the proper time $\tau$. Substituting Eqs. (\ref{eq:1.79}) and (\ref{eq:1.80}) into Eq.(\ref{eq:1.76}), we get

\begin{equation}
E^2=\dot{r}^2+V_{eff},
\label{eq:1.82}
\end{equation}

\noindent where

\begin{equation}
V_{eff}=\left(1-a-\frac{2mr^2}{r^3+2l^2m}\right)\left(\frac{J^2}{r^2}+L\right)
\label{eq:1.83}
\end{equation}

\noindent represents the effective potential for the geodesic motion in the space-time of a Hayward black hole with a cloud of strings.

Using the relation

\begin{equation}
\frac{dr}{dt}\frac{dt}{d\tau}=\frac{dr}{d\tau}\Rightarrow \left(\frac{dr}{dt}\right)^2\left(\frac{dt}{d\tau}\right)^2=\left(\frac{dr}{d\tau}\right)^2\Rightarrow\left(\frac{dr}{dt}\right)^2\dot{t}^2=\dot{r}^2
\label{eq:1.85}
\end{equation}

\noindent into Eq.(\ref{eq:1.82}) and using Eqs.(\ref{eq:1.83}) and eq.(\ref{eq:1.79}), we get

\begin{equation}
\left(\frac{dr}{dt}\right)^2=g(r)^2\left[1-\frac{g(r)}{E^2}\left(\frac{J^2}{r^2}+L\right)\right].
\label{eq:1.86}
\end{equation}
%%%%%%%%%%%%%%%%%%%%%%%%%%%%%%%%%%%%%%%%%%%%%%%%%%%%%%%%%%%
\subsubsection{Radial movement of a photon}

For the radial movement $(J=0)$ of a photon $(L=0)$, Eq.(\ref{eq:1.86}) can be written as

\begin{equation}
\left(\frac{dr}{dt}\right)^2=g(r)^2.
\label{eq:1.87}
\end{equation}

Substituting Eq.(\ref{eq:1.58}) into Eq.(\ref{eq:1.87}), we get the relation between the coordinates $t$ and $r$, which is given by

\begin{equation}
\pm t=\int\frac{1}{1-a-\frac{2mr^2}{r^3+2l^2m}}dr.
\label{eq:1.88}
\end{equation}

Using the Eq.(\ref{eq:1.82}), we can obtain the relation between the coordinate $r$ the proper time $\tau$ for the radial movement of a photon, which is given by 

\begin{equation*}
\left(\frac{dr}{d\tau}\right)^2=E^2,
\end{equation*}

\begin{equation}
\pm\tau=\frac{r}{E}.
\label{eq:1.89}
\end{equation}
%%%%%%%%%%%%%%%%%%%%%%%%%%%%%%%%%%%%%%%%%%%%%%%%%%%%%%%%%%
\subsubsection{Radial movement of a massive particle}

Now, let us consider the movement of massive particles  $(L=1)$ in radial trajectories $(J=0)$ nearby the black hole. From Eq.(\ref{eq:1.86}), we obtain

\begin{equation}
\left(\frac{dr}{dt}\right)^2=g(r)^2-\frac{g(r)^3}{E^2}.
\label{eq:1.90}
\end{equation}

Substituting Eq.(\ref{eq:1.58}) into Eq.(\ref{eq:1.90}), we can find the relationship between the coordinates $t$ and  $r$ for the radial movement of the particle:

\begin{equation}
\pm t=\int\frac{dr}{\sqrt{\left(1-a-\frac{2mr^2}{r^3+2l^2m}\right)^2-\frac{\left(1-a-\frac{2mr^2}{r^3+2l^2m}\right)^3}{E^2}}}.
\label{eq:1.91}
\end{equation}

From Eq.(\ref{eq:1.82}), we get the relationship between the proper time $\tau$ and the radial coordinate $r$:

\begin{equation*}
\left(\frac{dr}{d\tau}\right)^2=E^2-g(r),
\end{equation*}

\begin{equation}
\pm\tau=\int\frac{dr}{\sqrt{E^2-\left(1-a-\frac{2mr^2}{r^3+2l^2m}\right)}}.
\label{eq:1.92}
\end{equation}

\subsection{Effective potential}

\begin{figure*}
\centering
    \includegraphics[scale=0.8]{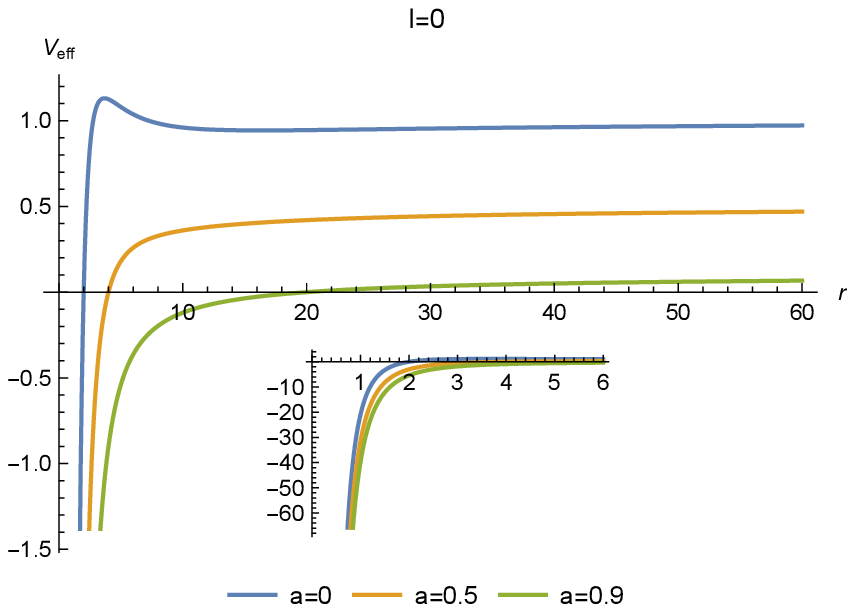}
    \includegraphics[scale=0.8]{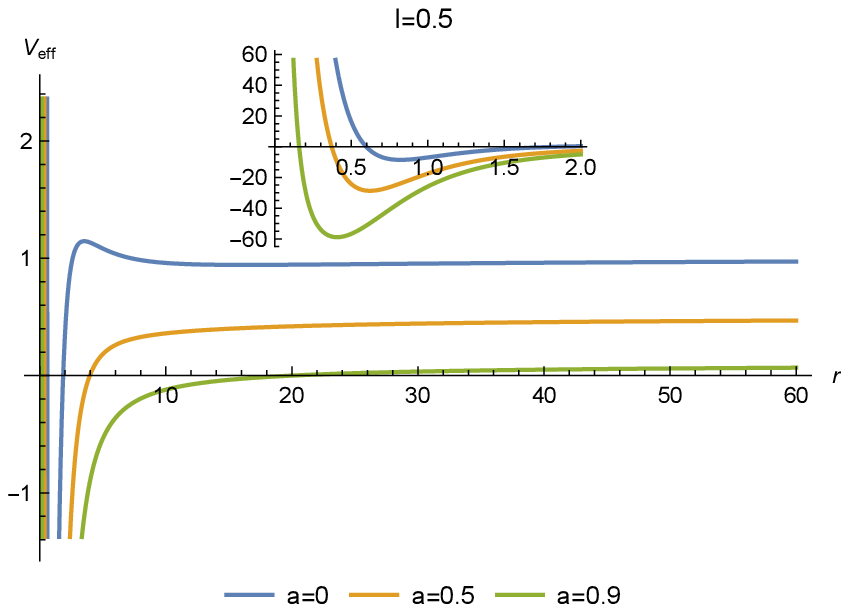}
    \includegraphics[scale=0.8]{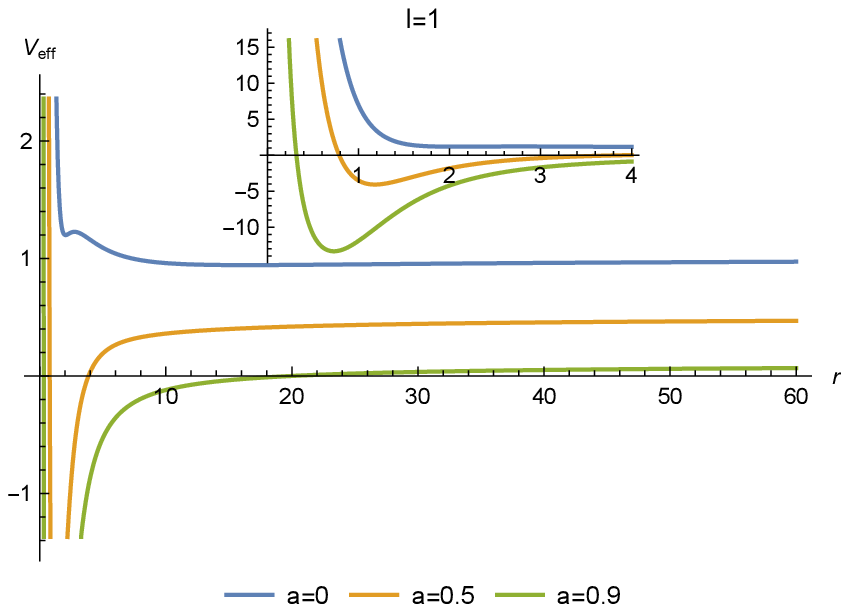}
    \includegraphics[scale=0.8]{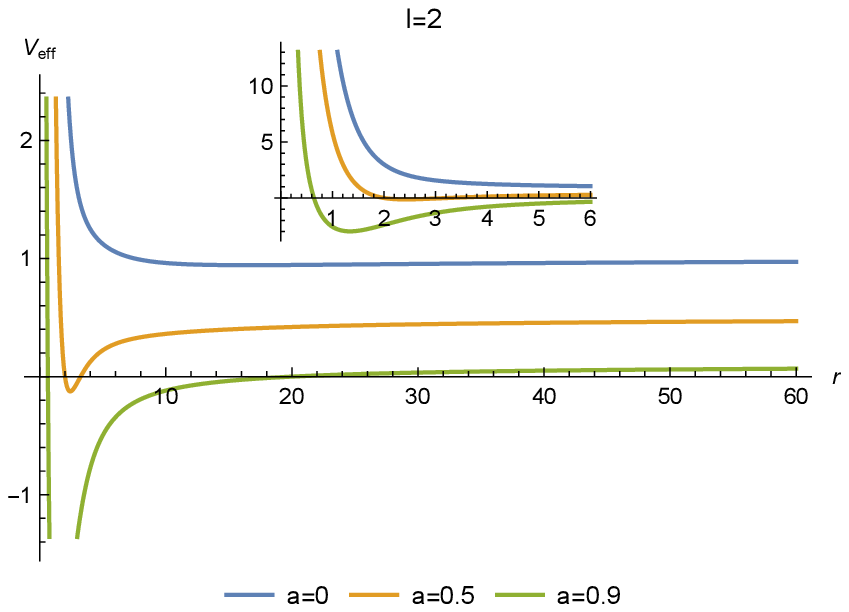}
    \caption{Effective potential for non-radial time-like geodesics ($L=1$ and $J^2=20$), for different values of $a$ and $l$.} 
\label{im5}
\end{figure*}

\begin{figure*}
\centering
\includegraphics[scale=0.8]{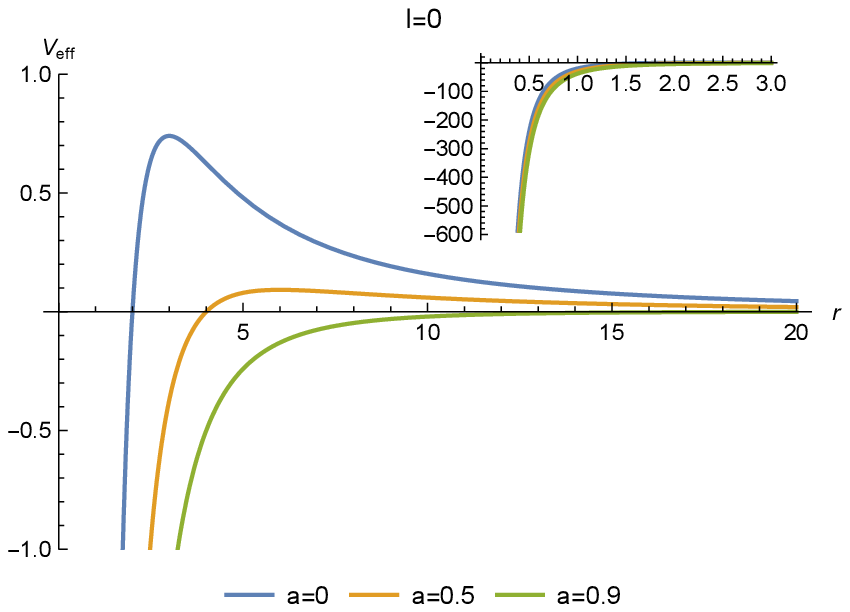}
\includegraphics[scale=0.8]{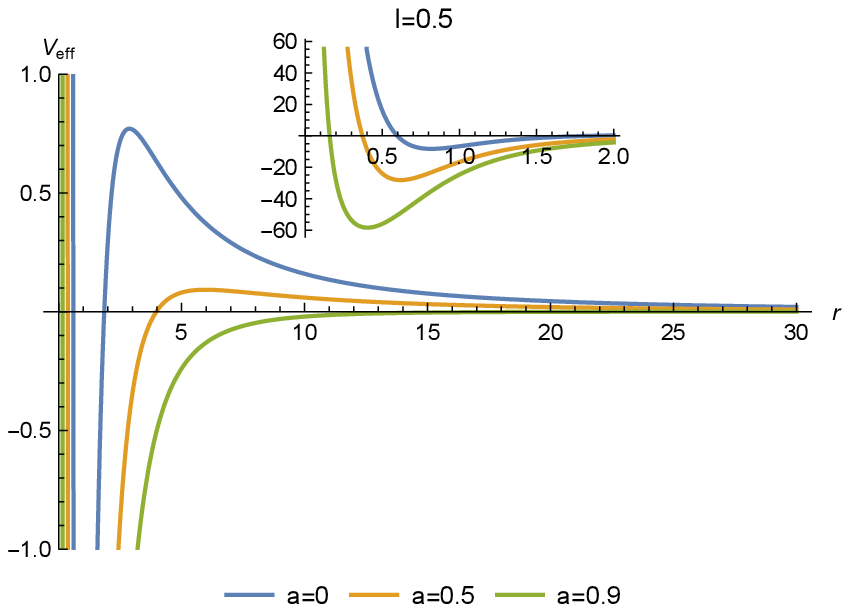}
\includegraphics[scale=0.8]{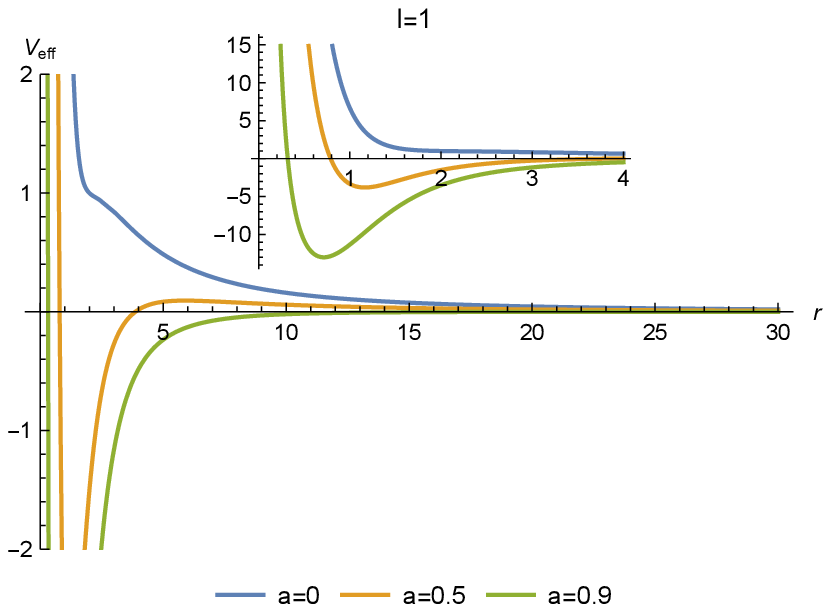}
\includegraphics[scale=0.8]{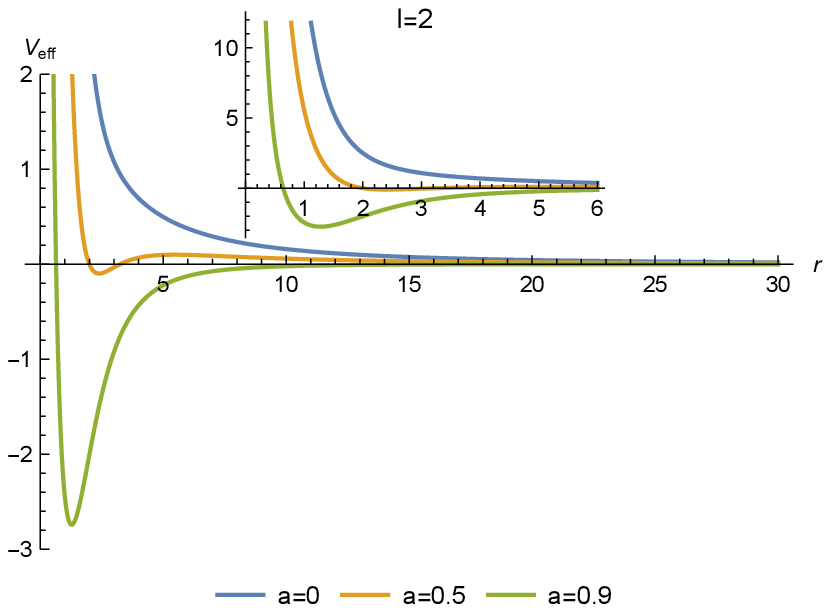}
\caption{Effective potential for non-radial null-like geodesics ($L=0$ and $J^2=20$), for different values of $a$ and $l$.} 
\label{im6}
\end{figure*}

\begin{figure*}
\centering
\includegraphics[scale=0.8]{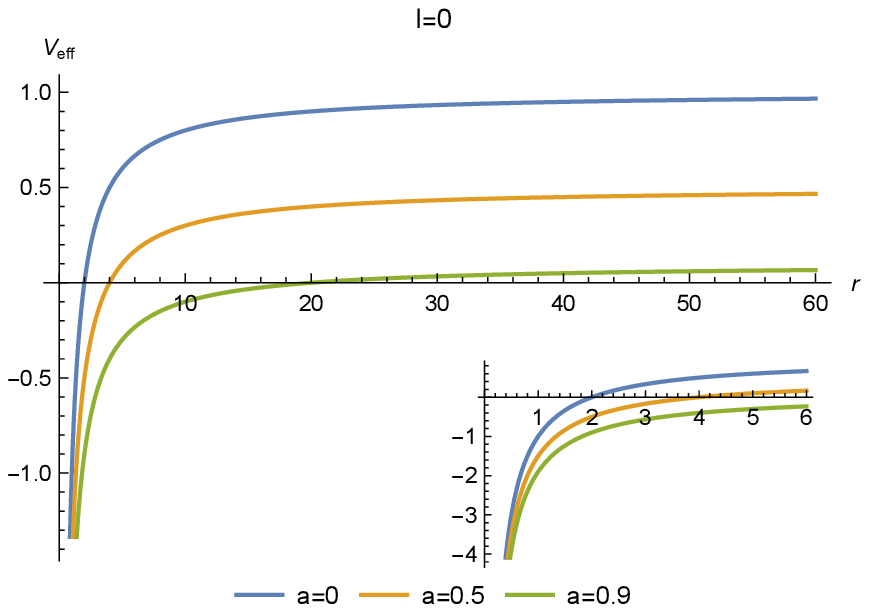}
\includegraphics[scale=0.8]{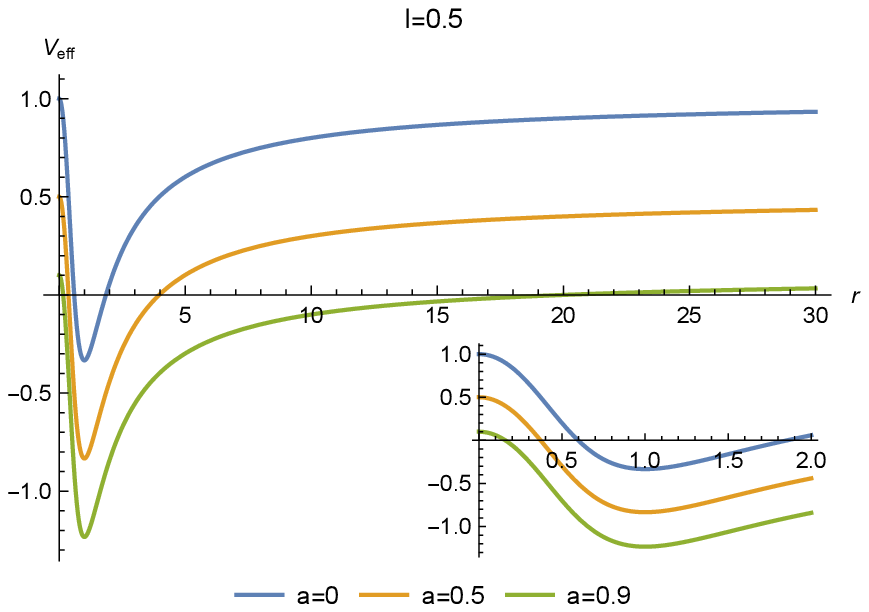}
\includegraphics[scale=0.8]{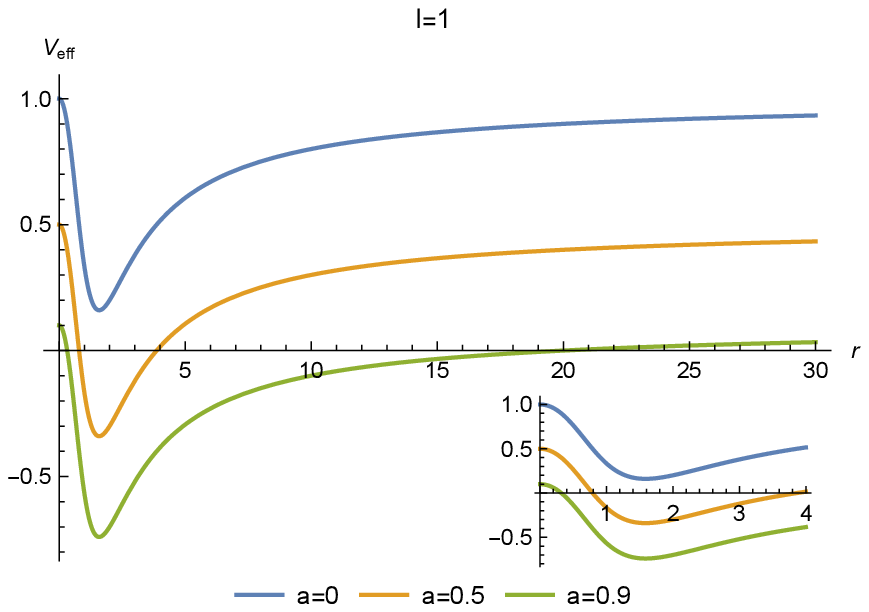}
\includegraphics[scale=0.8]{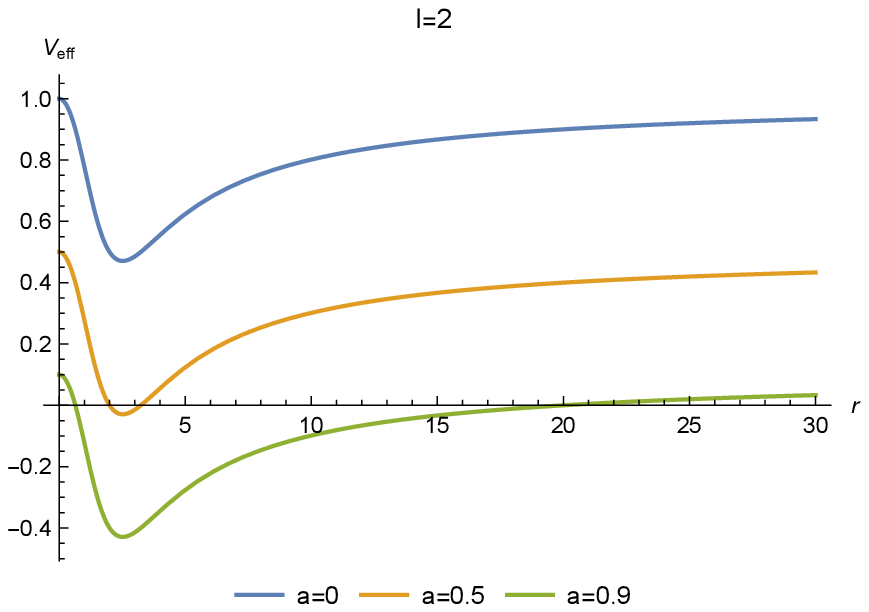}
\caption{Effective potential for radial time-like geodesics   ($J^2=0$ and $L=1$), for different values of $a$ and $l$.} 
\label{im7}
\end{figure*}

\begin{figure}
\centering
\includegraphics[scale=0.8]{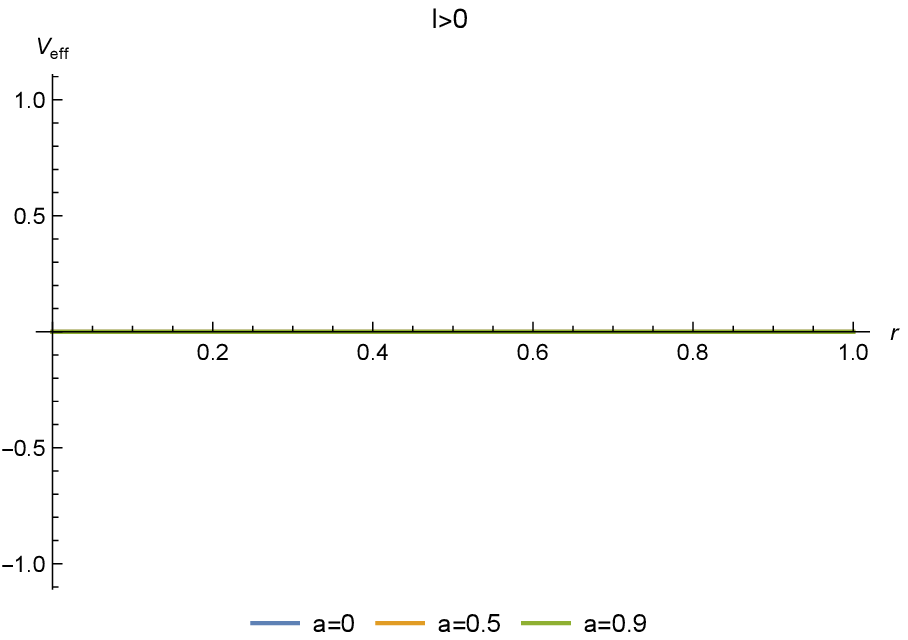}
\caption{Effective potential for radial null-like geodesics  ($J^2=0$ and $L=0$), for different values of $a$ and $l$.} 
\label{im8}
\end{figure}

The behavior of the effective potential ($V_{eff}$) of the geodesic motion, given by Eq. (\ref{eq:1.83}), can tell us about the behavior of a massive particle or a photon near the black hole. So, in Figs. \ref{im5} to \ref{im8}, we plot the effective potentials for different values of $a$ and $l$ for time-like and null-like geodesics. In some figures, we represent, in detail, $V_{eff}$ near the black hole ($r$ near zero).

In Fig. \ref{im5}, we represent the effective potential for geodesics ($L=1$ and $J^2=20$). We can observe that, for $l=0$, there is no stable circular geodesics, since the graphics do not show a local minimum for any value of $a$. On the other hand, for $l>0$, we can observe the possibility of the existence of stable circular geodesics, depending on the value of the cloud of strings parameter $a$. 

% \subsubsection{Potencial efetivo para geodésicas tipo-tempo não-radiais}

% \subsubsection{Potencial efetivo para geodésicas tipo-nulo não-radiais}

For non-radial time-like geodesics (Fig. \ref{im6}), we can observe that, in all cases, $V_{eff} \rightarrow 0$ for regions far from the black hole, $r \rightarrow \infty$. For $l=0$, there is no stable circular geodesics, since the graphics do not show local minima. The existence of stable circular orbits of photons around the black hole depends also on the cloud of strings parameter, as can be seen in Fig. \ref{im6}.

% \subsubsection{Potencial efetivo para geodésicas radiais tipo-tempo}

In which concerns the behavior of radial time-like geodesics, Fig. \ref{im7} shows us that the stability of radial movement does not occur for $l=0$. On the other hand, for $l>0$, there will always be stable geodesics.

% \subsubsection{Potencial efetivo para geodésicas radiais tipo-nulo}

Finally, we can observe in Fig. \ref{im8} that, for radial null-like geodesics  ($J^2=0$ and $L=0$), the effective potential is constant and equal to zero.

 %%%%%%%%%%%%%%%%%%%%%%%%%%%%%%%%%%%%%%%%%%%%%%%%%%%%%%%%%%%%%%
\section{Black Hole thermodynamics}
\label{sec4}

In this section, we study the thermodynamics of the Hayward black hole with a cloud of strings by examining the behavior of the mass, Hawking temperature, and heat capacity as a function of entropy.

%%%%%%%%%%%%%%%%%%%%%%%%%%%%%%%%%%%%%%%%%%%%%%%%%%%%%%%%%%%%%%
\subsection{Black hole mass}

Let $r_h$ be the radius of the horizon, thus we have that $g(r_h)=0$, where $g(r)$ is given by Eq.(\ref{eq:1.58}). Thus, we can write the mass of the black hole in terms of $r_h$ through the following equation:

\begin{equation}
m=\frac{(1-a)r_h^3}{2(r_h^2+(a-1)l^2)},
\label{eq:1.59}
\end{equation}

\noindent which is written in terms of the parameter that takes care of the presence of the cloud of strings, namely, $a$. Note that if $a=0$, we recover the mass of the regular Hayward black hole, without the cloud of strings, in terms of the horizon radius. By setting $l=0$ and $a=0$ we recover the mass of the Schwarzschild black hole. 

The area of the horizon can be calculated by

\begin{equation}
A=\int \sqrt{-g}d\theta d\phi=4\pi r_h^{2}.
\label{eq:1.60}
\end{equation}

On the other hand, the entropy of the black hole can be calculated through the area law \cite{bekenstein1973black}, using the relation

\begin{equation}
S=\frac{A}{4}=\pi r_h^2.
\label{eq:1.61}
\end{equation}

Thus, we can write the mass parameter as a function of the entropy as 

\begin{equation}
m=\frac{(1-a)S^{3/2}}{2\pi^{3/2}(\frac{S}{\pi}+(a-1)l^2)}.
\label{eq:1.62}
\end{equation}

\begin{figure*}
\centering
\begin{minipage}[!]{0.45\linewidth}
\includegraphics[scale=0.8]{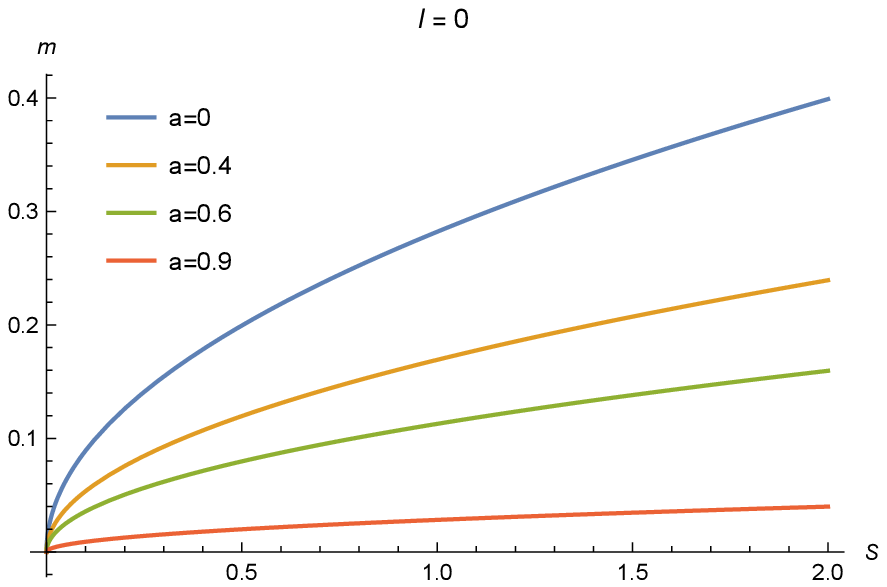}
\end{minipage}
\begin{minipage}[!]{0.45\linewidth}
\includegraphics[scale=0.8]{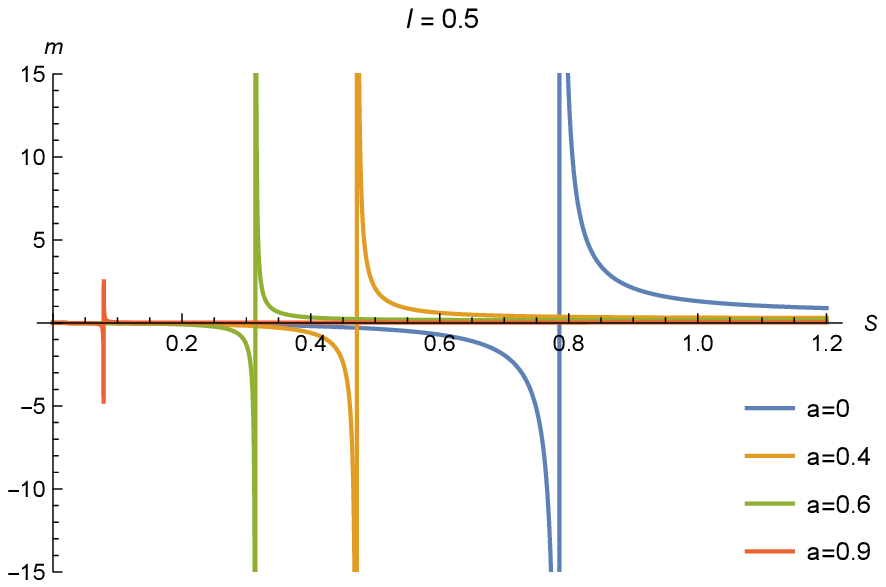}
\end{minipage}
\begin{minipage}[!]{0.45\linewidth}
\includegraphics[scale=0.8]{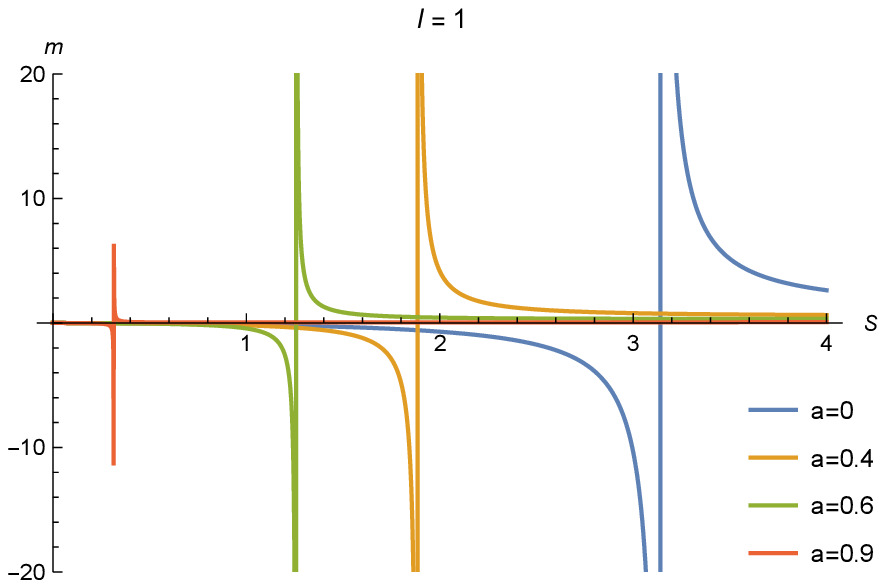}
\end{minipage}
\begin{minipage}[!]{0.45\linewidth}
\includegraphics[scale=0.8]{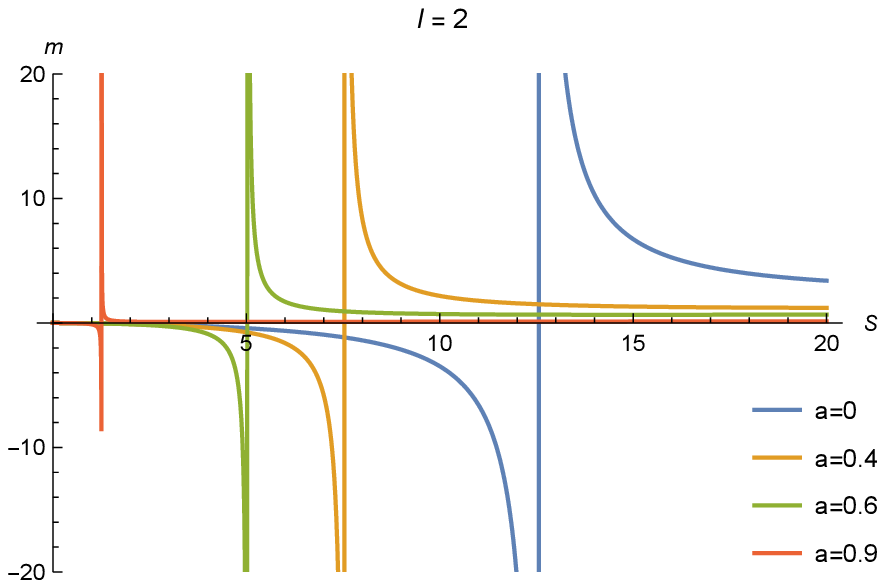}
\end{minipage}
\caption{Black hole mass as a function of the entropy $m(S)$ for different values of $a$ and $l$.} 
\label{im2}
\end{figure*}

In Figure \ref{im2}, we represent the behavior of the mass parameter, $m$, as a function of the entropy of the black hole, $S$, in different situations. Note that, for the Schwarzschild 
black hole, ($a=0$ and $l=0$), the mass parameter presents only positive values for positive values of the entropy,$S$. Similar behavior is obtained when  ($l=0$), thus we return to the Schwarzschild black hole scenario, but now, with a cloud of strings, $0<a<1$. If we consider the Hayward black hole, it is possible to notice that the mass parameter has positive and negative values depending on the parameters of the black hole. This is also repeated when we consider the cloud of strings in Hayward's black hole space-time. It is important to note that the cloud of string parameter modifies the phase transition point for Hayward black hole surrounded by this cloud.
 
%%%%%%%%%%%%%%%%%%%%%%%%%%%%%%%%%%%%%%%%%%%%%%%%%%%%%%%%%%%%%%
\subsection{Hawking temperature}

The surface gravity $(\kappa)$ for the Hayward black hole with a cloud of strings can be calculated using the following expression:

\begin{equation}
\kappa=\frac{g'(r)}{2}\left|\frac{}{}_{r_{h}}\right.,
\label{eq:1.63}
\end{equation}

\noindent with $'$ denoting the derivative with respect to the radial coordinate. Hawking showed that the black hole emits radiation and its corresponding temperature, Hawking temperature, for stationary space-time, is given by \cite{hawking1975particle}:

\begin{equation}
T_\kappa=\frac{\kappa}{2\pi}.
\label{eq:1.64}
\end{equation}

It is worth remembering that for vacuum space-time with spherical symmetry, the first law of thermodynamics gives

\begin{equation}
dm=T_f dS\rightarrow T_f=\frac{dm}{dS},
\label{eq:1.65}
\end{equation}

\noindent where $m$ and $S$ are the total energy and entropy of the system, respectively, and $T_f$ is the temperature of the black hole predicted in the first law. Taking into account the area law given by Eq.(\ref{eq:1.61}), the first law of thermodynamics does not provide a correct way to calculate the temperature of regular black holes \cite{ma2014corrected,maluf2018thermodynamics}. 

Thus, the Hawking temperature, $T_\kappa$ and $T_f$, for the Hayward black hole obtained through equations Eqs. (\ref{eq:1.64}) and (\ref{eq:1.65}), respectively, are not equivalent, that is, the first law, given by Eq.(\ref{eq:1.65}), is not appropriate to calculate the correct temperature for regular black holes since such black holes are not vacuum solutions.

Using Eq. (\ref{eq:1.64}) and with $\kappa$ given by (\ref{eq:1.63}), it is possible to calculate the Hawking temperature, $T_\kappa=T$, for the Hayward black hole with the cloud of strings:

\begin{equation}
T=\frac{m r_h(r_h^3-4l^2m)}{2\pi(r_h^3+2l^2m)^2}.
\label{eq:1.66}
\end{equation}

Putting Eq.(\ref{eq:1.59}) into Eq.(\ref{eq:1.66}), we obtain the following result

\begin{equation}
T=\frac{(1-a)[r_h^2+3(a-1)l^2]}{4\pi r_h^3}.
\label{eq:1.67}
\end{equation}

Substituting $r_h=(\frac{S}{\pi})^{1/2}$ in Eq.(\ref{eq:1.67}), we finally obtain the Hawking temperature expression as a function of entropy for the Hayward black hole with the cloud of strings:

\begin{equation}
T=\frac{(1-a)\sqrt{\pi}[3(a-1)l^2+\frac{S}{\pi}]}{4S^{3/2}}.
\label{eq:1.68}
\end{equation}

\begin{figure*}
\centering
\begin{minipage}[!]{0.45\linewidth}
\includegraphics[scale=0.8]{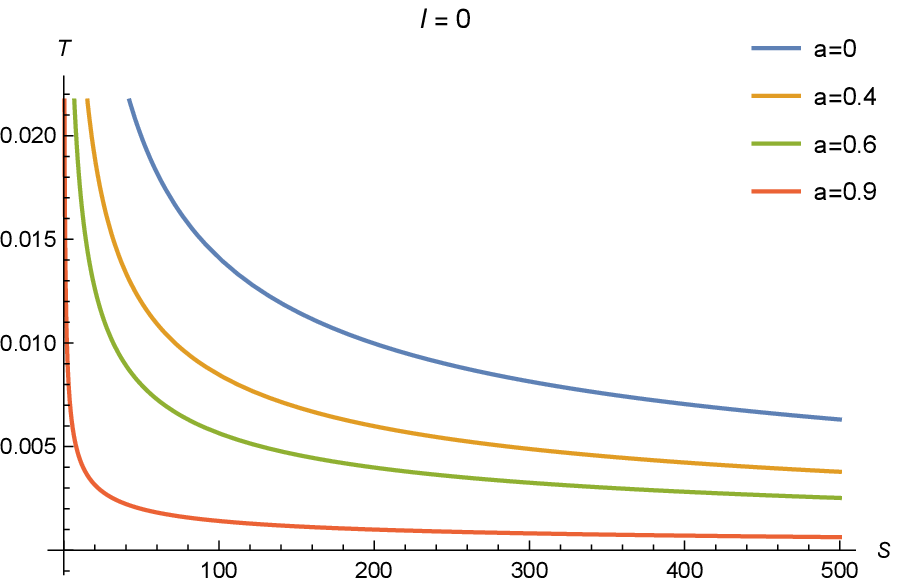}
\end{minipage}
\begin{minipage}[!]{0.45\linewidth}
\includegraphics[scale=0.8]{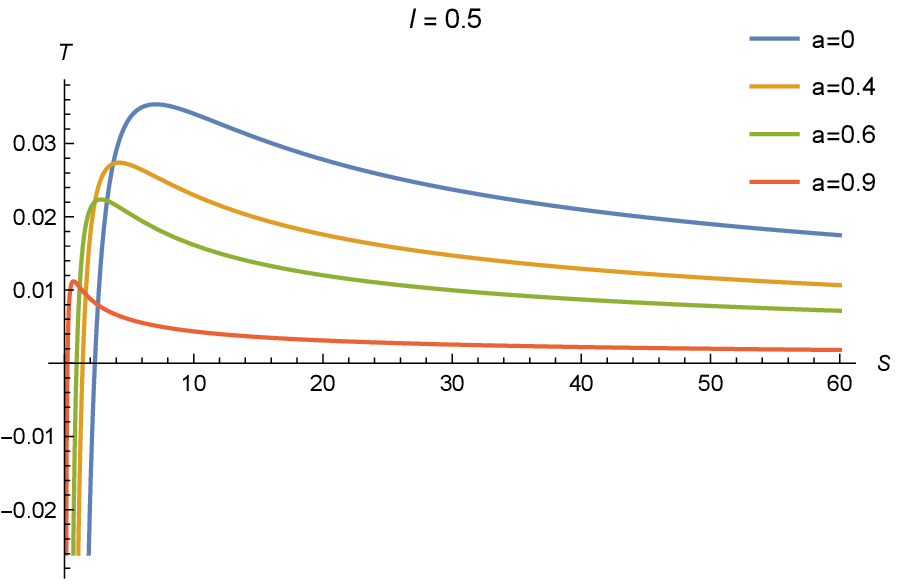}
\end{minipage}
\begin{minipage}[!]{0.45\linewidth}
\includegraphics[scale=0.8]{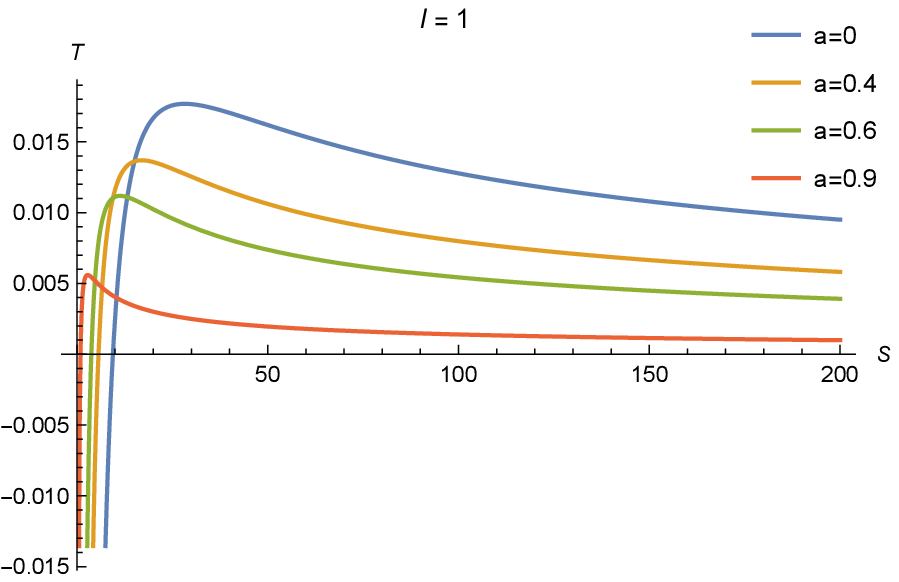}
\end{minipage}
\begin{minipage}[!]{0.45\linewidth}
\includegraphics[scale=0.8]{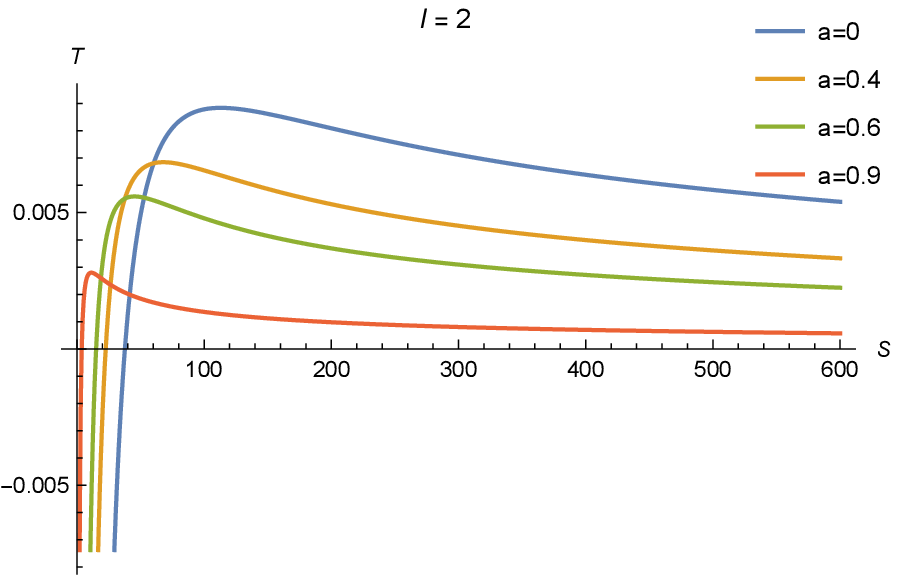}
\end{minipage}
\caption{Black hole temperature as a function of entropy  $T(S)$ for different values of $a$ and $l$.} 
\label{im3}
\end{figure*}

In figure \ref{im3}, we represent the behavior of the temperature parameter, $T$, as a function of the entropy of the black hole, $S$, in different situations. Note that for the Schwarzschild space-time ($a=0$ and $l=0$) the temperature parameter presents only positive values for $S>0$. Analogously, for the Letelier space-time ($l=0$ and $0<a<1$). When we consider the Hayward space-time ($l=1$ and $a=0$), it is already possible to notice that the temperature parameter will present positive and negative values depending on the black hole parameters. This is also repeated when we consider the cloud of strings in Hayward spacetime ($l=1$ and $0<a<1$).
%%%%%%%%%%%%%%%%%%%%%%%%%%%%%%%%%%%%%%%%%%%%%%%%%%%%%%%%%%%%%%
\subsection{Heat capacity}

Heat capacity provides information about the thermodynamic stability of a system. We can calculate the heat capacity of the Hayward black hole with the cloud of strings from the following expression:

\begin{equation}
C=T\frac{\partial S}{\partial T}=T\left(\frac{\partial T}{\partial S}\right)^{-1}.
\label{eq:1.69}
\end{equation}

Substituting Eq.(\ref{eq:1.68}) into Eq.(\ref{eq:1.69}), we find the following expression for the heat capacity as a function of the black hole entropy:

\begin{equation}
C=-\frac{2S[3(a-1)l^2\pi+S]}{9(a-1)l^2\pi+S}.
\label{eq:1.70}
\end{equation}

\begin{figure*}
\centering
\begin{minipage}[!]{0.45\linewidth}
\includegraphics[scale=0.8]{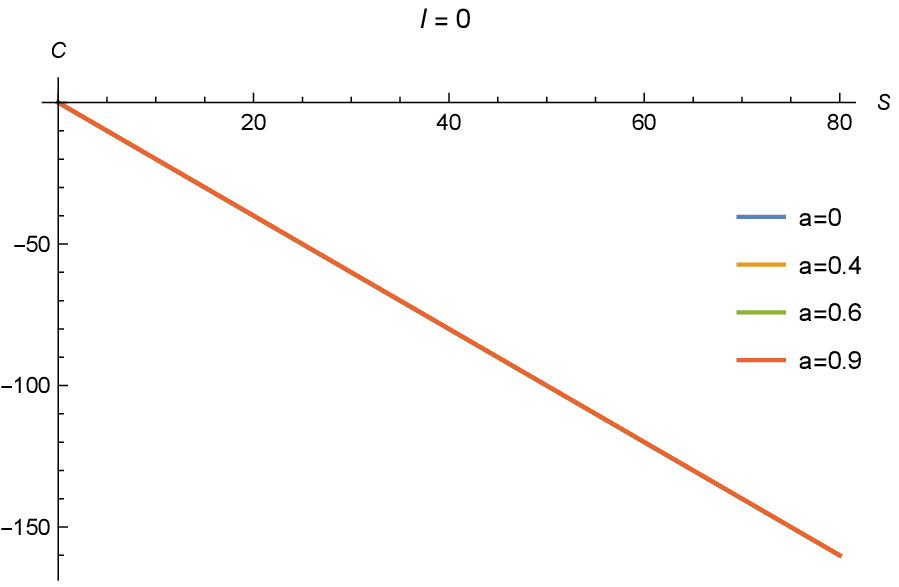}
\end{minipage}
\begin{minipage}[!]{0.45\linewidth}
\includegraphics[scale=0.8]{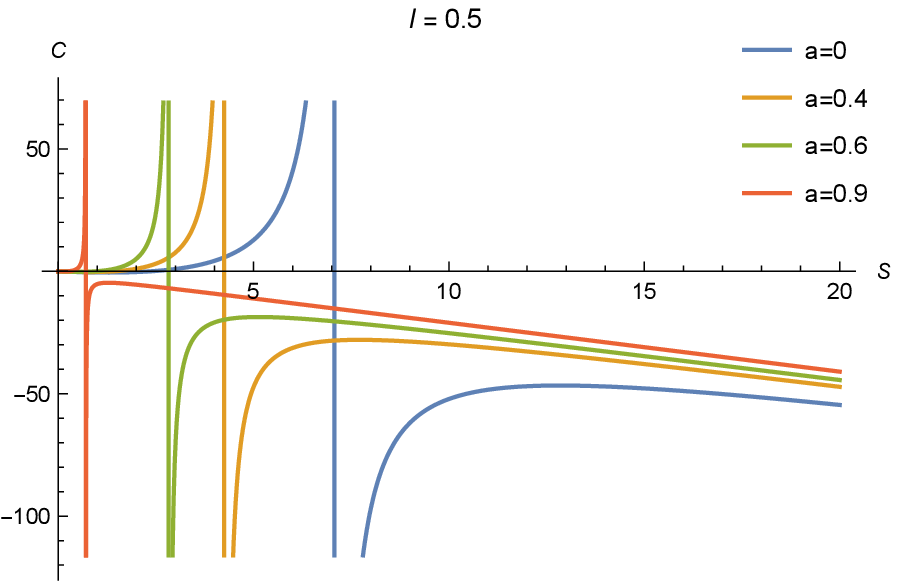}
\end{minipage}
\begin{minipage}[!]{0.45\linewidth}
\includegraphics[scale=0.8]{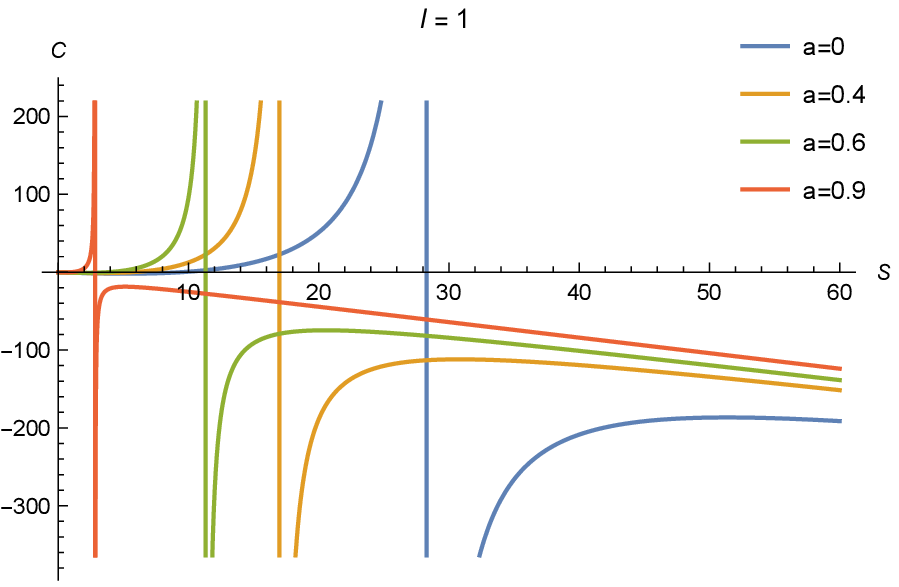}
\end{minipage}
\begin{minipage}[!]{0.45\linewidth}
\includegraphics[scale=0.8]{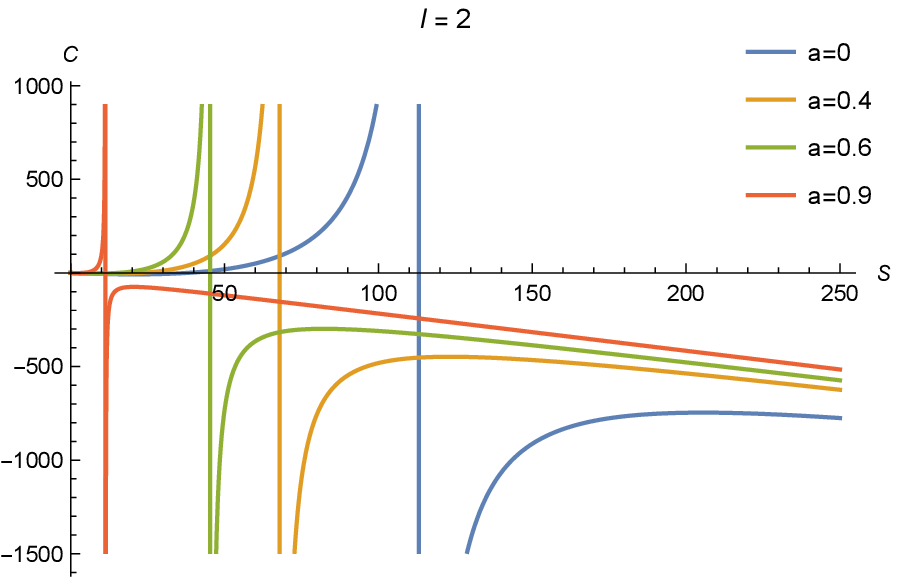}
\end{minipage}
\caption{Heat capacity as a function of the entropy $C(S)$ for different values of $a$ and $l$.} 
\label{im4}
\end{figure*}

The behavior of the heat capacity as a function of  entropy, for different values of $l$ and $a$, is given in Figure \ref{im4}. For $l=0$, Eq.(\ref{eq:1.70}) reduces to $C = -2S$,  as expected for the case of Schwarzschild spacetime. In this case, the heat capacity is negative for $S>0$, which indicates an unstable thermodynamic system.

When we take into account Hayward ($l=1$,$a=0$) and the Hayward space-time with the cloud of strings ($l=1$,$0<a<1$), there are values for the entropy for which the heat capacity
acquires positive values. This means that the Hayward black hole can be unstable or stable depending on the values of the entropy considered. It is important to point out that the string parameter
plays an important role in the behavior of the heat capacity, modifying the phase transition point. Thus, unlike the Schwarzschild black hole, which is thermodynamically unstable, the Hayward black hole and the Hayward black hole with a cloud of strings present regions of stability, depending on the chosen parameters.

%
%%%%%%%%%%%%%%%%%%%%%%%%%%%%%%%%%%%%%%%%%%%%%%%%%
%

%
%%%%%%%%%%% Conclusão %%%%%%%%%%%%%%%%%%%%
%

\section{Concluding remarks}
\label{sec5}

In conclusion, we can say that the cloud of strings has an interesting role in different aspects related to the Hayward black hole surrounded by this cloud.

Firstly, let us analyze the Kretschmann scalar in the limit $r\rightarrow 0$, whose result is $\infty$. This means that the inclusion of the cloud of strings turns the black hole singular. In other words, the metric of the Hayward black hole with a cloud of strings is singular at the origin.

With relation to the possible roots of $g(r)$, and considering a critical mass, $m_*$,
given by Eq(\ref{eq:1.58.1}), three different scenarios are present, namely: 
(i) the black hole mass is higher than the critical mass, $m>m_*$, thus $g(r)$ has two real roots;
(ii) the black hole mass is equal to the critical mass, $m=m_*$, then, $g(r)$ has a unique real root, which is equal to $r_*$; and (iii)   $m<m_*$, in which case, $g(r)$ has no real roots.

The effective potential ($V_{eff}$) of the geodesic motion, given by Eq. (\ref{eq:1.83}), is represented in Figs. \ref{im5} to \ref{im8}, for different values of $a$ and $l$ for time-like and null-like geodesics. As shown, in some cases, in which $l=0$, there are no stable circular geodesics, irrespective of the values of $a$. On the other hand, for $l>0$, we can observe the possibility of the existence of stable circular geodesics, depending on the value of the cloud of strings parameter $a$. 

For non-radial time-like geodesics (Fig. \ref{im6}), we can observe that, in all cases, $V_{eff} \rightarrow 0$ for regions far from the black hole, $r \rightarrow \infty$. For $l=0$, there is no stable circular geodesics. Otherwise, there are stable circular orbits of photons around the black hole, depending on the presence of the cloud of strings, as can be seen in Fig. \ref{im6}.

In which concerns the behavior of radial time-like geodesics, shown in Fig. \ref{im7}, for $l>0$, there exist stable geodesics in all situations.

The behavior of the mass parameter, m, as a function of the entropy of the black
hole, S, in different situations, are very similar, as we can see in Fig. \ref{im2}.
It is worth calling attention to the fact that preserves the shape, irrespective of the values of the parameter that codifies the presence of the cloud of strings. The only difference is a shift to the left, as the values of this parameter increase.

When the Schwarzschild black hole space-time (a = 0 and l = 0) is taken into account, the
temperature parameter presents only positive values for
$S > 0$. The same behavior occurs when a cloud of strings is added  to Schwarzschild black hole space-time, namely, $l = 0$
and $0 < a < 1$. Now, if the Hayward black hole is considered, $l = 1$ and $a = 0$, the temperature will assume positive
and negative values depending on the black hole 
parameters. The same situation appears when the presence of a cloud of string
is considered in Hayward.

The heat capacity
assumes positive values or negative ones, depending on the values of entropy, when the cloud of strings is present. This means that the Hayward black hole with a cloud of strings can be unstable or stable depending on
depending on the values of the entropy considered. This behavior is analogous to the one obtained when the cloud is absent. Otherwise, the role of the cloud of strings is to shift the graphs to the left.

It is important to point out that the cloud of strings parameter
plays an important role in the behavior of the heat capacity, modifying the points of the phase transitions, as well as on the Hawking temperature, black hole mass, effective  potential, geodesics and horizons, and removing the regular character of the Hayward regular black hole solution, by removing it and turning the solution singular.

%%%%%%%%%%%%%%%%%%%%%%%%%%%%%%%%%%%%%%%%%%%%%%%%%%%%%%%%

\begin{acknowledgements}

V.B. Bezerra is partially
supported by CNPq-Brazil ( Conselho Nacional de Desenvolvimento Científico e Tecnológico) through Research Project No. 307211/2020-7.

F. F. Nascimento and J. M. Toledo acknowledge Departamento de Fisica, Universidade Federal da Paraiba, for hospitality.

\end{acknowledgements}

%BibTeX users please use one of
% \bibliographystyle{spmpsci}      % mathematics and physical sciences
\bibliographystyle{spphys}       % APS-like style for physics
\bibliography{refs.bib}   % name your BibTeX data base

% % Non-BibTeX users please use
% \begin{thebibliography}{}
% %
% % and use \bibitem to create references. Consult the Instructions
% % for authors for reference list style.
% % Format for Journal Reference
% Author, Article title, Journal, Volume, page numbers (year)
% Author, Book title, page numbers. Publisher, place (year)
% \end{thebibliography}

\end{document}